\documentclass{epl}

\title{Comparative study between two quantum spin systems 
KCuCl$_{3}$ and TlCuCl$_{3}$ }

\author{ T. Saha-Dasgupta\inst{1} and Roser Valent\'\i\inst{2}} 

\institute{
\inst{1} S.N. Bose National Centre for Basic Sciences, 
JD Block, Sector 3,
 Salt Lake City, Kolkata 700098, India.\\
\inst{2} Fakult\"at 7, Theoretische Physik,
 University of the Saarland,
66041 Saarbr\"ucken, Germany.}


\begin{document}

\maketitle

\begin{abstract}
We have performed an {\it ab initio} study
of the electronic structure of two isostructural
 quantum spin systems, KCuCl$_{3}$ and TlCuCl$_{3}$,
which have recently attracted much attention  due to 
their unconventional magnetic properties. Our first-principles analysis shows
unambiguously the role of Tl, as opposed to structural differences
between the two compounds, in making TlCuCl$_{3}$ a {\it strongly} coupled
 s=$\frac{1}{2}$ dimer 
system compared to KCuCl$_{3}$ which shows a {\it weakly}
coupled  s=$\frac{1}{2}$ dimer behavior. 
Good agreement with the existing analysis of  
inelastic neutron scattering results has been observed.

\end{abstract}



{\it Introduction.-} The behavior of low-dimensional quantum spin systems is
  dominated by 
zero-point spin fluctuations.  As a consequence, they show a variety
 of  interesting  magnetic properties which are not present in their
 classical counterparts. In the last years a great amount of effort,
 both theoretically and experimentally, has been dedicated to
 the study of these quantum spin systems.  An important example
 of such systems are materials
having a singlet ground state with a finite energy 
 gap in the spin excitation spectra like
even-leg ladder compounds\cite{lad}, s= $\frac{1}{2}$
 alternating chain compounds
\cite{ac}, spin-Peierls \cite{sp} or 
spin-dimer compounds \cite{dimer}. 
 In particular, the family of coupled spin-dimer systems
 ACuCl$_3$ (A = K, Tl) is being currently intensively studied in connection to
 its magnetic properties.   KCuCl$_{3}$  has
   a spin excitation gap $\Delta$ $\sim$ 2.6 meV 
 and the  analysis of inelastic neutron scattering
 (INS)\cite{cavadini_00,kcucl} data indicate that this system is 
 a spin-dimer system  with {\it weak} three-dimensional interdimer
couplings.  TlCuCl$_3$,   though being isostructural to 
 KCuCl$_{3}$,  has
   a spin excitation gap  $\Delta$ $\sim$ 0.65meV  about four times
smaller than that of  KCuCl$_{3}$ and a  saturation field 
twice as large as that for KCuCl$_3$. Accordingly, 
the behavior of this system
 has been proposed\cite{INS2,INS1}
 to be that of a three-dimensional {\it strongly} coupled
 spin-dimer system. The small size of the spin gap makes these 
 compounds very suitable for studying the effects of the application of
 an external magnetic field on the quantum properties of the system.
 Recently \cite{mag},
 magnetic field-induced N\'eel ordering  has been observed in 
 TlCuCl$_3$ for magnetic fields $H$ higher than the
  gap field $H_g$ = $\Delta /g \mu_B$ $\sim$ 6 T 
and has been interpreted
 as a Bose-Einstein condensation
 of  excited triplets (magnons) \cite{BEC}. New nuclear magnetic
resonance (NMR) investigations\cite{Vyaselev_02}
 have also shown that this
 phase transition is accompanied by changes in the crystal
lattice parameters,
 what is an indication of an important spin-phonon coupling in this
 material. The observation of field-induced phase transition
 has been also reported in KCuCl$_3$ for a magnetic field
$H > H_g$ $\sim$ 22 T \cite{Oosawa_01}.
 Magnetization measurements\cite{dis} were also performed in 
 the mixed spin system (Tl$_{1-x}$K$_x$)CuCl$_3$ where the $x$ substitution
 creates bond randomness in the system and the field-induced
phase transition was studied as a function of the K concentration.
  Motivated by the great amount of exotic phenomena 
 observed in these materials, we have carried out first principles study
 of the electronic properties of these systems
 in order to give a microscopic foundation 
 to their behavior and analyze
 the similarities and differences between the two systems.

\begin{figure}
\includegraphics[width=8cm]{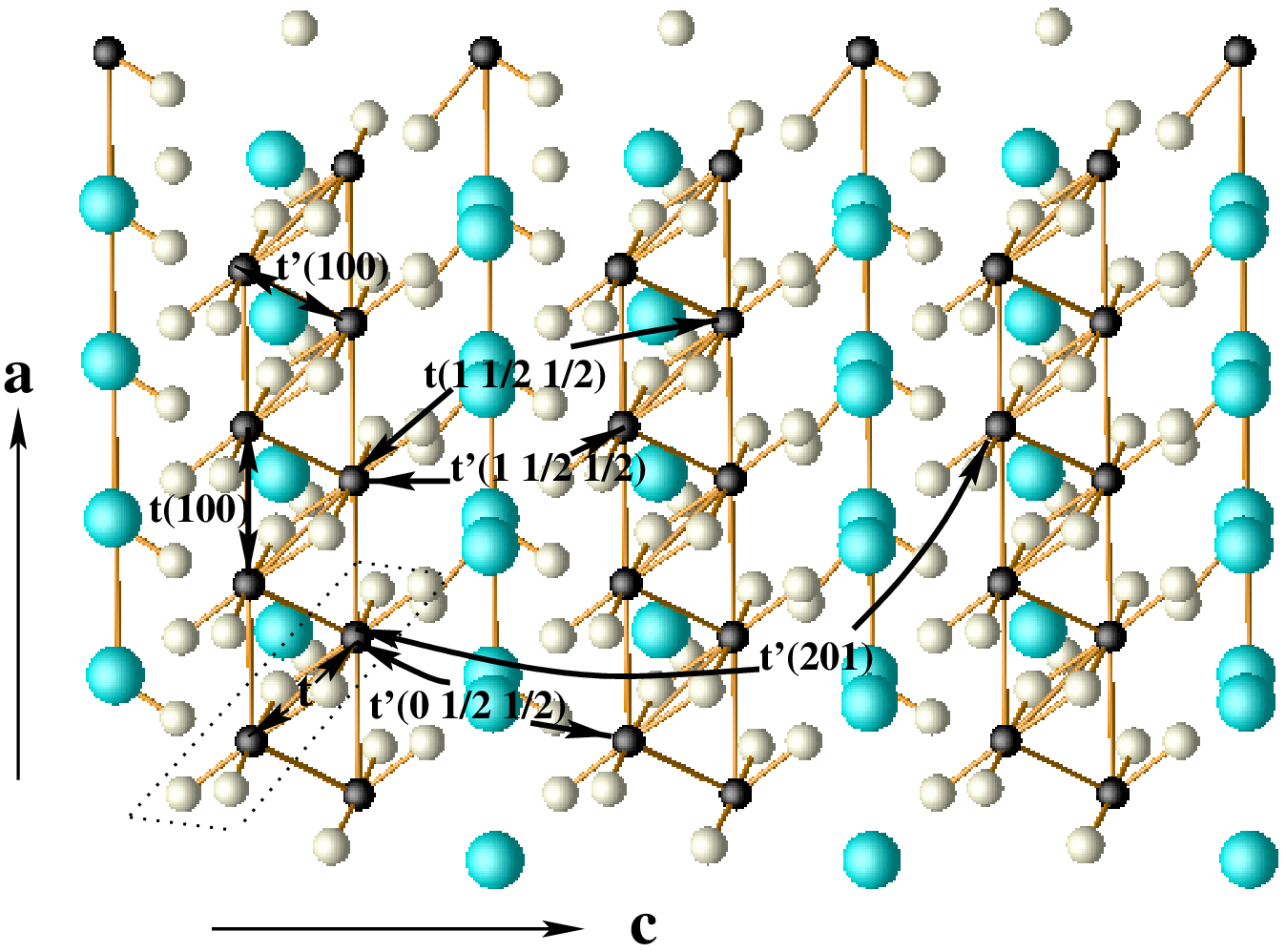}
 \caption{.
 }
 \label{structure}
\end{figure}

{\it Crystal Structure.-} Both KCuCl$_{3}$ and TlCuCl$_{3}$ crystallize  in the monoclinic $P2_{1}/c$
space group\cite{struc} with four formula units per unit cell.
 The lattice parameters
for KCuCl$_{3}$ are given by $a = 4.029 \AA$, $b = 13.785 \AA$, $c= 8.736 \AA$
and $\beta = 97.20^{o}$, while those for TlCuCl$_{3}$ are $a = 3.982 \AA$, 
$b = 14.144 \AA$, $c= 8.890 \AA$ and $\beta = 96.32^{o}$. Therefore, compared
to KCuCl$_{3}$, the TlCuCl$_{3}$ lattice is somewhat compressed along
the $a$-axis and enlarged  in the $bc$ plane.  The degree of monoclinicity
also varies by some extent. Both structures are formed by edge-sharing 
CuCl$_{6}$ octahedra which build zig-zag chains running along the $a$-axis
(fig. 1).  These chains are located at the corners and
 center of the unit cell in the
$bc$ 
plane. K$^{+}$/Tl$^{+}$ ions are distributed in between these chains.
 The CuCl$_{6}$ octahedra are highly distorted due to 
the Jahn-Teller active
Cu$^{2+}$ ion with large elongation along one direction resulting into
nearly planar Cu$_{2}$Cl$_{6}$ dimer clusters  which are the basic
unities in the structure (marked with the box in fig. \ref{structure}).

\begin{figure}
\includegraphics[width=16cm]{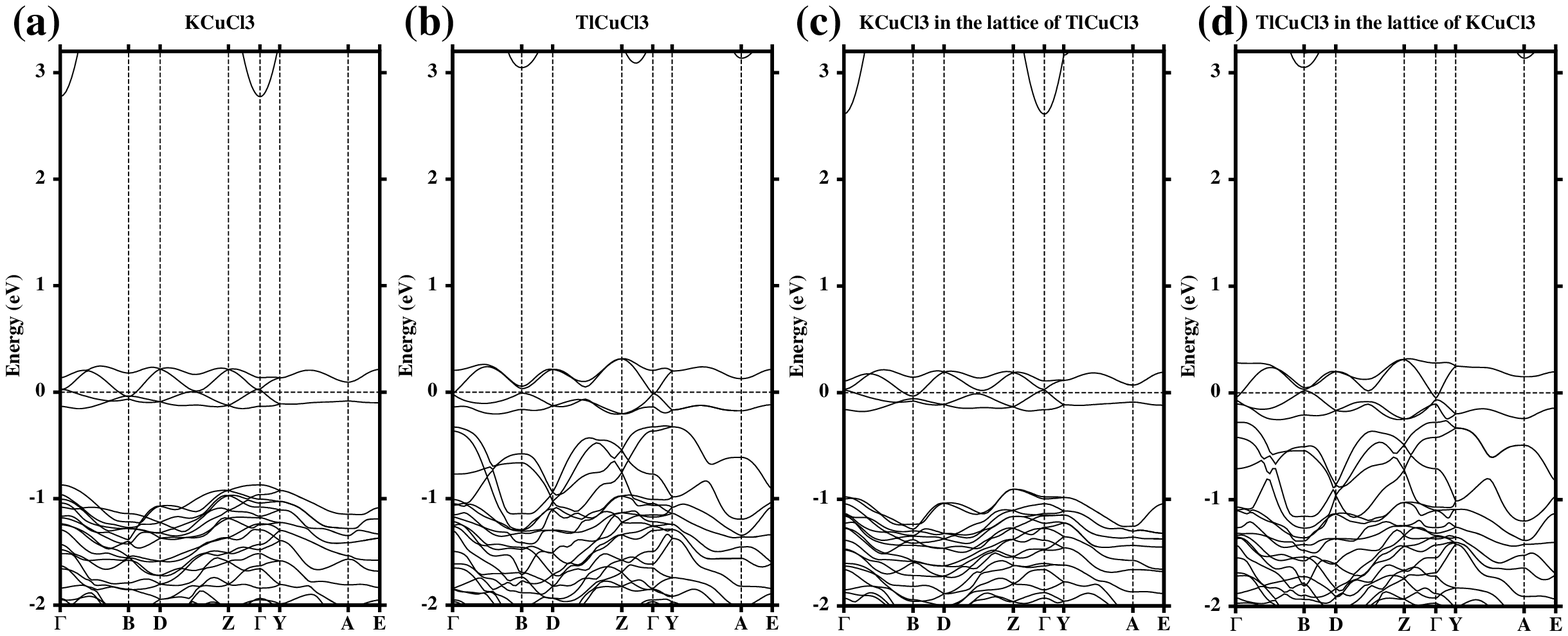}
 \caption{  }
 \label{bands}
\end{figure}

{\it Band Structure.-} 
In panels (a) and (b) of fig.\ \ref{bands}
 we show the band-structures of KCuCl$_{3}$ and 
TlCuCl$_{3}$ along various symmetry directions. First-principles Density
 Functional Theory (DFT)
calculations have been carried out within the framework of the 
state-of-art full-potential
 linearized augmented plane wave (LAPW)
method\cite{lapw} and the
 linearized muffin tin orbital (LMTO) method\cite{lmto} based on
the local-density-approximation (LDA). The 
band-structures obtained by the two methods are in overall
 agreement with each other.
The predominant features of the band structure are the complex of four 
narrow bands close to the Fermi level formed by Cu d$_{x^2-y^2}$
 orbitals (in the local frame of reference) contributed
by each Cu atom in the unit cell, admixed with Cl p-states. 
These  bands are half-filled and the insulating groundstate observed
 in these compounds should be explained by the effect of strong 
correlations which are not fully considered in the LDA approximation. 
 This set of bands is separated from the low-lying valence bands by a gap of
 about 0.3 to 0.5 eV  \cite{explanation} and from the high-lying
  excited
  bands, which are
dominated by states from the K/Tl atoms, by another 
gap of about 3 eV. Concentrating on the low-energy bands close to
the Fermi-level,  we see that the bands are dispersive along all the symmetry
directions suggesting the compounds to be three-dimensionally
 coupled systems in
agreement with the findings of INS measurements
 \cite{cavadini_00,kcucl,INS2,INS1}.
 Since the dispersion behavior 
is similar between KCuCl$_{3}$ and TlCuCl$_{3}$,  
the predominant interaction pathways should be  alike for both 
compounds. 
 The total bandwidth of the four-band complex close to the 
Fermi level
is somewhat larger for TlCuCl$_{3}$ ($\approx$ 0.5 eV)  compared to that
of KCuCl$_{3}$ ($\approx$ 0.4 eV) suggesting the intradimer coupling 
to be stronger  in
TlCuCl$_{3}$ than in
 KCuCl$_{3}$. 
 We also observe here that the
 bands in TlCuCl$_{3}$ are in general more dispersive  than in   KCuCl$_{3}$
 what indicates larger interdimer interactions in  TlCuCl$_{3}$
 compared to KCuCl$_{3}$ \cite{explanation_2}.

 This change in the
strength of the intra- and interdimer couplings between the two compounds
 could be caused by (i)
the changes in the structural parameters
between the two compounds or by (ii) the role of the Tl$^{+}$ ion compared to
that of the K$^{+}$ ion. In order to have an  understanding of which of these
 effects
 determines predominantly the behavior of these systems,
 we have carried out model calculations of  KCuCl$_{3}$
 by considering the  structural parameters  of the Tl-compound and
 viceversa, we  computed TlCuCl$_{3}$
 by considering the structural parameters  of the K-compound.
 The corresponding
band structures are shown in panels (c) and (d) of fig.\ \ref{bands}.  
 As we see, changing
the lattice parameters has a minor effect on the band structure (compare
fig.\ \ref{bands}(a) and fig.\ \ref{bands}(c))
 while substituting K$^{+}$ by Tl$^{+}$ in the
same lattice has a significant effect making the band dispersions comparable
to that of the Tl-compound.
 This exercise already points unambiguously to the role
of the Tl$^+$
 ion in enhancing the strength of the coupling as compared
to that of changes in the structural parameters. In the next section
 we will make
this analysis more quantitative in terms of various hopping
integrals.

{\it Low-energy Hamiltonians: Hopping integrals.-}
We have used the LMTO-based downfolding method, which has been proposed, 
implemented\cite{newlmto} and applied to a number of cases\cite{example} in
recent years, in order to obtain the low-energy effective Hamiltonians
that describe the behavior of a system.
 This method of deriving the low-energy Hamiltonians by 
integrating out (downfolding)
 the high energy degrees of freedom results in  
Hamiltonians defined in the basis of effective orbitals. The process takes
into account the proper effective contribution
from the orbitals that are being downfolded.
For the present compounds, we have derived the low-energy Hamiltonians
defined in the basis of effective Cu d$_{x^2-y^2}$ orbitals, by keeping only 
the d$_{x^{2}-y^{2}}$ orbital for each Cu atom in the unit cell 
and integrating
out all the rest. The Fourier transform of this
 few-orbital downfolded Hamiltonian
provides the various hopping integrals, $t_{ij}$,
 between these effective orbitals and the corresponding tight-binding (TB)
Hamiltonian can be written as
$H = \sum_{(i,j)} t_{ij} (c_i^{\dagger} c_j + h.c.)$
where $i$ and $j$ denote a pair of Cu$^{2+}$ ions.
These hopping integrals thus, form the first-principles set of parameters
obtained without any fitting procedure containing the fingerprint of the
pathways involved in the hopping processes.

Table I shows the various hopping integrals for KCuCl$_{3}$, 
TlCuCl$_{3}$ and the two model compounds  KCuCl$_{3}$ considering the 
 lattice parameters of
TlCuCl$_{3}$ and TlCuCl$_{3}$ in the lattice of KCuCl$_{3}$. The various
hopping pathways are marked in fig. 1. 
The hoppings are named according to the
 various magnetic couplings shown in fig. 1 of ref.  \cite{INS1} i.e.
 $t(l\ m\ n)$
 ($t'(l\ m\ n)$)
 denote the hopping parameters between two equivalent (nonequivalent)
sites  (in terms of the corresponding spins for magnetic
interaction) in dimers separated by a
 lattice vector $\vec{r} = l a \hat{x} + m b \hat{y} + n c \hat{z}$.  In 
Table I we also show the magnetic couplings obtained in various references
by fitting the experimental  INS data.

Although,  due to the complexity of the exchange
 pathways, there are no
simple relationships connecting the hopping and magnetic
 exchange integrals,
the relative strength of the hopping parameters in the
 K- and Tl-compound can be easily noticed to be
in good qualitative agreement with the relative strength
shown by the magnetic exchange parameters.

 Considering first the intradimer coupling, 
 the distortion in the nearest neighbor Cu-Cl-Cu dimer
bond (see the marked box in fig. 1)
from 90$^o$ towards a linear bonding of 180$^o$ angle increases the 
importance of the antiferromagnetic (AF) coupling over the  ferromagnetic
(FM) coupling. The Cu-Cl-Cu bond angles in these compounds are
 $\approx$ 96$^{o}$ which is larger
than 94$^{o}$, the  {\it upper bound} angle  given  in the
edge-shared cuprates\cite{af+fm}
 for a 
possible ferromagnetic coupling, thereby providing the intradimer
 exchange coupling J
  of AF nature.  Considering the superexchange mechanism to be
the effective mechanism of exchange in the dimer, and therefore
relating the magnetic exchange integral J with the hopping
 integral $t$ through J=$4t^2/U$ where $U$ is the effective on-site
Coulomb repulsion, we observe that the ratio
t$^{2}_{kcucl_3}$/t$^{2}_{tlcucl_3}$ $\sim$  0.75 obtained from the
 {\it ab initio} analysis compares very well with the ratio of
J's obtained by fitting the INS data, 0.76 in ref. \cite{INS1} 
and 0.79 in ref. \cite{INS2}.

Focusing now on the interdimer interactions,
note that
 the values of the hopping matrix elements in 
 TlCuCl$_{3}$ are larger than those for  KCuCl$_{3}$, thus concluding 
 that TlCuCl$_{3}$ will be a strongly coupled dimer system compared to  
KCuCl$_{3}$ which can
be characterized as a weakly coupled dimer system in agreement with
the conclusions drawn from INS data and already conjectured from
band dispersions. Quantitatively, the largest enhancement of the
interdimer coupling between TlCuCl$_{3}$ and KCuCl$_{3}$ occurs
for neighboring dimers in the  (1,0,-2) plane, i.e.
 $(l\ m\ n) = (2\ 0\ 1)$. The other
dominant interdimer hoppings are $t(1\ 1/2\ 1/2)$ and $t(1\ -1/2\
1/2)$\cite{explanation_3}, the last one
being 
significantly enhanced in the TlCuCl$_{3}$ system compared to that of
KCuCl$_{3}$. The LDA band dispersions of the four-band complex 
are reproduced reasonably well by considering the intradimer hopping
matrix element
and the following interdimer hoppings $t^{'}(2\ 0\ 1)$ and $t(1\ 1/2\ 1/2)$ for
KCuCl$_{3}$ and the additional hopping of $t(1\ -1/2\ 1/2)$
for TlCuCl$_{3}$.

 The analysis of the two model calculations (panel (c) and (d) of
 fig.\ \ref{bands}) shows
 the evident role of the Tl$^+$ ion  in enhancing
the hopping matrix elements between the Cu$^{2+}$ ions. Comparing the 2nd and 5th column
of Table I, the hopping matrix elements turn out to be similar
between KCuCl$_{3}$ and the model system KCuCl$_{3}$ in the lattice
of TlCuCl$_{3}$, while keeping the lattice parameters fixed
and substituting K by Tl, the hoppings are affected 
significantly (4th column) with larger intra-  and interdimer
hoppings \footnote{Considering the model calculation of TlCuCl$_{3}$ in
the lattice of KCuCl$_{3}$, the lattice shrinks along the $b-$axis and
expands along the $a-$axis (by a factor 8 smaller
 than the shrinkage along the $b-$axis)
 which is a distorsion similar to the recently
observed changes in the crystal lattice when the system undergoes
a phase transition to a N\'eel state induced by the application
of an external magnetic field\cite{Vyaselev_02}. Although the relative
change in the $b-$parameter for this model calculation is much larger
than that reported for the field-induced transition, this study may
provide indications of the changes in the couplings caused by this
strain effect. }.  In order to quantify the action of Tl$^+$, we
show 
in fig. 3  the Tl and K partial density
of states (PDOS) in TlCuCl$_{3}$ and  KCuCl$_{3}$ respectively. As one sees,
the contribution of Tl in the four narrow low energy bands is much
larger compared to that of K which implies larger hybridization of
Tl with the Cu d$_{x^2-y^2}$ orbitals in TlCuCl$_{3}$ compared to that
of K in KCuCl$_{3}$. This relatively large hybridization effect
 from the Tl atoms
situated in between the double chains   can be understood in
 terms of the proximity of 
the Cu d$_{x^2-y^2}$ and the Tl  6$s$, 6$p$ energy states as well as the
extended nature of these orbitals in comparison to K 4$s$.  This behavior
 leads to a
strongly coupled 
network of Cu-dimers in TlCuCl$_{3}$,  while   KCuCl$_{3}$
 behaves as a collection of Cu-dimers
 weakly interacting in three-dimensions. This description is
 also supported by electron density analysis of the LDA results.
Interestingly, this conclusion is opposite to the case of CaV$_{2}$O$_{5}$ and
MgV$_{2}$O$_{5}$ where the subtlety in structural differences turn
out to be responsible for the differences in the 
behavior of the two systems\cite{cavo+mgvo}.

{\it Conclusions.-}
  We have presented a comparative {\it ab initio} study of the electronic
 properties of KCuCl$_3$ and TlCuCl$_3$, two isostructural
 quantum spin systems which have been the subject of recent interest
 due to their unconventional magnetic properties. 
 It has been the purpose of this study: (i) to obtain a microscopic
description
of the properties of these systems and its comparison with the behavior
 predicted from INS. 
 (ii) to understand  the  microscopic origin of the different behavior
      between  these two compounds.

By means of the downfolding-TB analysis which reduces the 
information provided by LDA to an effective low-energy Hamiltonian
in tight-binding basis,
we could succesfully corroborate the coupled dimer behavior predicted
for these systems by INS. In agreement with the predictions from
the INS data analysis, we found in KCuCl$_3$ the interdimer couplings
 to be  weak, while TlCuCl$_3$ manifests itself as a strongly coupled
dimer system.
We also showed that  the origin of the differences in the interdimer
coupling between the two systems is related to
 the relative contribution of Tl compared to that of K to 
 the chemical bonding between two Cu$^{2+}$ ions.
Our study is also indicative of possible changes thay may occur
in the coupling constants due to lattice changes caused by strain
effects\cite{Vyaselev_02}.

\acknowledgments

We acknowledge useful discussions with Peter Lemmens and H. Tanaka
 and R.V. thanks
the Deutsche Forschungsgemeinschaft for finantial support through
 a Heisenberg fellowship.



\newpage

{\bf FIGURE CAPTIONS} \\ \\

\noindent
{1. Structure of KCuCl$_3$ (TlCuCl$_{3}$). Small black and white atoms
represent Cu and Cl respectively while the large gray atoms
represent K(Tl). The middle double chain is shifted with respect
to other two double chains along $b-$direction by half a lattice
constant. Various hoppings are marked (see text for details).}\\ \\

\noindent
{2. Band structure of KCuCl$_{3}$ (panel (a)), TlCuCl$_{3}$
(panel (b)), KCuCl$_{3}$ in the lattice of TlCuCl$_{3}$ (panel(c))
and TlCuCl$_{3}$ in the lattice of KCuCl$_{3}$ (panel(d)) along
the symmetry directions $\Gamma$ = (0 0 0),  B=(-$\pi$,0,0),
D=(-$\pi$,0,$\pi$),
Z=(0,0,$\pi$), $\Gamma$,
Y=(0,$\pi$,0), A = (-$\pi$,$\pi$,0),
E=(-$\pi$,$\pi$,$\pi$).
}\\ \\

\noindent
{3. Tl(solid) and K(dotted) contributions to
the density of states of TlCuCl$_{3}$ and KCuCl$_{3}$ respectively. 
}\\

{\bf TABLE CAPTIONS} \\ \\

{\it I. Hopping integrals obtained from our ab initio analysis
for TlCuCl$_{3}$,
 KCuCl$_{3}$ and the two model systems considered in the text and magnetic interactions from ref.\ \cite{INS2} and ref.\ \cite{INS1}
 for TlCuCl$_{3}$ and
 KCuCl$_{3}$.
 The negative sign denotes a ferromagnetic coupling.}

\newpage


\begin{figure}
\includegraphics[width=8cm]{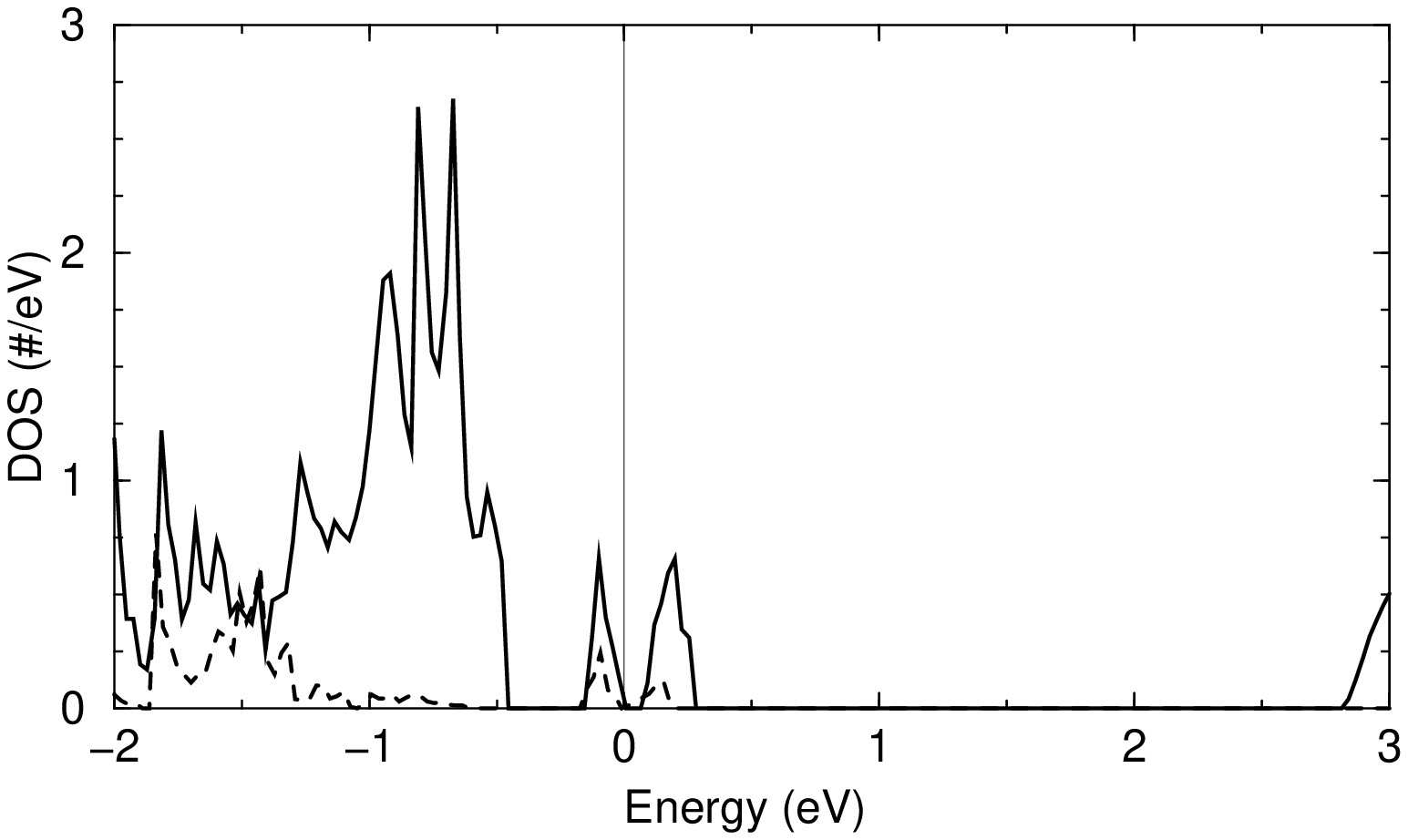}
 \caption{  }
 \label{dos}
\end{figure}

\begin{table}
\caption{  }
\begin{tabular}{|c|c|c|c|c|}
t's and J's & KCuCl$_{3}$ & TlCuCl$_{3}$ & TlCuCl$_{3}$ in & KCuCl$_{3}$ in \\ 
in meV &  &  & KCuCl$_{3}$
& TlCuCl$_{3}$ \\ 
& & & lattice & lattice \\ \hline
t & 116 & 134 & 144 & 117 \\
J & 4.34 ref\cite{INS1} & 5.68 ref\cite{INS1} & & \\
 &  4.29 ref\cite{INS2} & 5.42 ref\cite{INS2} & & \\\hline
t(100) & 2.72 & 6.81 & 4.08 & 1.00 \\
t'(100) & 9.52 & 21.76 & 24.49 & 5.40 \\
J$_{eff}$(100)= & -0.21 ref\cite{INS1} & -0.46 ref\cite{INS1} & & \\
1/2[2J(100) - J'(100)] &  -0.21 ref\cite{INS2} & -0.47 ref\cite{INS2} & & \\\hline
t'(201) & 36.73 & 65.31 & 72.11 & 35.37 \\
J$_{eff}$(201) = & -0.45 ref\cite{INS1} & -1.53 ref\cite{INS1} & & \\
-1/2J'(201) &  -0.34 ref\cite{INS2} & -1.43 ref\cite{INS2} & & \\\hline
t(1 1/2 1/2) & 40.82 & 36.73 & 42.17 & 35.40 \\
t(1 -1/2 1/2) & 4.08 & 32.18 & 40.50 & 2.72 \\
t'(1 1/2 1/2) & 5.19 & 12.24 & 10.88 & 8.16 \\
J$_{eff}$(1 1/2 1/2)= & 0.28 ref\cite{INS1} & 0.49 ref\cite{INS1} & & \\
1/2[J(1 1/2 1/2) - J'(1 1/2 1/2)] &  0.37 ref\cite{INS2} & 0.62 ref\cite{INS2} & & \\\hline
t(0 1/2 1/2) & 5.44 & 8.16 & 6.80 & 8.01 \\
t'(0 1/2 1/2) & 14.97 & 19.05 & 16.33 & 17.63 \\
J$_{eff}$(0 1/2 1/2)= & -0.003 ref\cite{INS1} & -0.06 ref\cite{INS1} & & \\
1/2[J(0 1/2 1/2) - J'(0 1/2 1/2)] &  &  & & \\\hline
\end{tabular}
\end{table}


\begin{thebibliography}{99}

\bibitem{lad} Dagotto E. and Rice T. M., Science {\bf 271}, (1996) 618.

\bibitem{ac} Garrett A. W., Nagler S. E., Tennant D. A.,
             Sales B. C. and Banes T., Phys. Rev. Lett {\bf 79} (1997)
             745.

\bibitem{sp} Hase M., Terasaki I. and Uchinokura K.,
	     Phys. Rev. Lett {\bf 70} (1993) 3651.

\bibitem{dimer} Tennant D. A., Nagler S. E., Garrett A. W., Barnes T.,
                Torardi C. C., Phys. Rev. Lett. {\bf 79}, 745 (1997).

\bibitem{cavadini_00} Cavadini N., Heigold G., Henggeler W., Furrer A.,
		G\"udel H. -U., Kr\"amer K., Mutka H.,
                J. Phys.:Condens. Matter {\bf 12} (2000) 5463.

\bibitem{kcucl} M\"uller M. and Mikesha H. -J., J. Phys.:
              Condens. Matter {\bf 12}, (2000) 7633.

\bibitem{INS2}  Cavadini N., Heigold G., Henggeler W., Furrer A.,
		G\"udel H. -U., Kr\"amer K., Mutka H., Phys. Rev
		B {\bf 63} (2001) 172414.



\bibitem{INS1}  Oosawa A., Kato T., Tanaka H., Kakurai K., 
                M\"uller M. and Mikesha H. -J., Phys. Rev. B
                {\bf 65} (2002) 094426.




\bibitem{mag} Oosawa A., Ishii M. and Tanaka H., J. Phys.:
              Condens. Matter {\bf 11}, (1999) 265.

\bibitem{BEC} T. Nikuni, M. Oshikawa, A. Oosawa and H. Tanaka,
              Phys. Rev. Lett {\bf 84}, 5868 (2000); M. Matsumoto, 
              B. Normand, T. M. Rice, and M. Sigrist, to appear in
              Phys. Rev. Lett (2002)

\bibitem{Vyaselev_02} Vyaselev O., Takigawa M., Vasiliev A.,
                      Oosawa A., and Tanaka H.
                      {\it to be published}.


\bibitem{Oosawa_01} Oosawa A., Tanaka H., Takamasu T.,
                    Abe H., Tsujii N., and Kido G.,
                    Physica B {\bf 294-295}, (2001) 34.

\bibitem{dis}    Oosawa A. and Tanaka H., {\it cond-mat/0203293}, to
                 appear in Phys. Rev. B (2002).

\bibitem{struc}  Willett R. D., Dwinggins C., Kruh R. F. and 
                 Rundle R. E., J. Chem. Phys.,{\bf 38}, (1963) 2429; 
                 Tanaka H., Oosawa A., Kato T., Uekusa H., Ohashi Y.,
                 Kakurai K. and Hoser A., J. Phys. Soc. Jpn {\bf 70}
                 (2001) 939. 

\bibitem{lapw}   Blaha P., Schwarz K. and Luitz J., WIEN97,
             {\it A Full Potential Linearized Augmented Plane 
        Wave Package for Calculating Crystal Properties}, (Karlheinz
         Schwarz, Techn. Univ. Wien, Vienna 1999). ISBN 3-9501031-0-4.
       [Updated version of Blaha P., Schwarz K., Sorantin P.,
         and Trickey S. B., Comp. Phys. Commun. {\bf 59}, (1990) 399].

\bibitem{lmto}  Andersen O. K., Phys. Rev. B {\bf 12}, (1975) 3060.

\bibitem{explanation} The gap between the occupied valence states
and the four-band complex at the Fermi level is underestimated in the  LMTO
method compared to the LAPW method, what can be attributed to the
atomic sphere approximation invoked in LMTO. This discrepancy
however does not affect the low-energy physics at the Fermi level
that we will be
concerned with.

\bibitem{explanation_2} This interpretation can be understood by the
 fact that the bandstructure of an isolated dimer system ({\it i.e.} 
 systems having only non-zero 
 intradimer interaction) is formed by two dispersionless bands, the separation
 of which is proportional to the strength of the intradimer
 interaction. Switching on the interactions between isolated dimers
 (inter-dimer interactions) brings in dispersion to the otherwise
 dispersionless band structure. Therefore, the bandwidth of the
 four-band complex in the present case gives us
 an estimate of the intradimer coupling-strength while the 
 degree of dispersion
 of these bands can be related to the interdimer couplings.

\bibitem{newlmto} Andersen O. K. and Saha-Dasgupta T.,
Phys. Rev. {\bf B62}, R16219 (2000) and references there in.

\bibitem{example} Pavarini E., Dasgupta I., Saha-Dasgupta T.,
Jepsen O.,
Andersen O. K., 
Phys. Rev. Lett. {\bf 87} (2001) 047003; Valent\'\i~ R.,
Saha-Dasgupta T., 
Alvarez J. V., Pozgajcic K., Gros C.,
Phys. Rev. Lett. {\bf 86}, (2001) 5381;  M\"uller T.,
Anisimov V.I., Rice T. M., Dasgupta I., Saha-Dasgupta T.,
Phys. Rev. {\bf B57}, (1998) R12655.


\bibitem{af+fm} Mizuno Y., Tohyama T., Maekawa S., Osafune T.,
Motoyama N., Eisaki H. and Uchida S., Phys. Rev. {\bf B57}, (1998) 5326 .



\bibitem{explanation_3} While the interdimer distances $t^{'}(1\
1/2\ 1/2)$ and $t^{'}(1\ -1/2\ 1/2)$, $t^{'}(0\
1/2\ 1/2)$ and $t^{'}(0\ -1/2\ 1/2)$ are all equivalent, $t(1\
1/2\ 1/2)$ and $t(1\ -1/2\ 1/2)$ are not equivalent since
they  involve different
pathways.



\bibitem{cavo+mgvo} Korotin M. A., 
Anisimov V. I., Saha-Dasgupta T. and Dasgupta I.,
J. Phys. (Condens. Matter) {\bf 12} (2000) 113.



\end{thebibliography}
\end{document}